\documentclass[journal,10pt]{IEEEtran}

\usepackage{cite}
\usepackage{graphicx}
\usepackage{array}
\usepackage{mdwmath}
\usepackage{amssymb}
\usepackage{mdwtab}
\usepackage{stfloats}
\usepackage[tight,footnotesize]{subfigure}
\usepackage{amsmath,amsthm}
\usepackage{threeparttable}
\usepackage{color}
\usepackage{url}
\usepackage{algpseudocode}
\usepackage{algorithm}
\usepackage{multicol}
\usepackage{epstopdf}
\usepackage{mathrsfs}
\usepackage{comment}

\algrenewcommand{\algorithmicrequire}{\textbf{Input:}}
\algrenewcommand{\algorithmicensure}{\textbf{Output:}}

\newtheorem{proposition}{Proposition}

\begin{document}

\title{\LARGE Movable Antenna Enhanced NOMA Short-Packet Transmission }

\author{Xinyuan~He,
        Wen~Chen,~\IEEEmembership{Senior~Member,~IEEE,}
        Qingqing~Wu,~\IEEEmembership{Senior~Member,~IEEE,}
        Xusheng~Zhu,
        and~Nan~Cheng,~\IEEEmembership{Senior~Member,~IEEE}% <-this % stops a space
\thanks{X. He, W. Chen, Q. Wu, and X. Zhu are with the Department of Electronic Engineering, Shanghai Jiao Tong University, Shanghai 200240, China (email: euphoria\_0428@sjtu.edu.cn; wenchen@sjtu.edu.cn; qingqingwu@sjtu.edu.cn; xushengzhu@sjtu.edu.cn).}% <-this % stops a space
\thanks{N. Cheng is with the School of Telecommunications Engineering, Xidian University, Xi’an 710071, China (e-mail: nancheng@xidian.edu.cn).}% <-this % stops a space

}

\maketitle

\begin{abstract}
This letter investigates a short-packet downlink transmission system using non-orthogonal multiple access (NOMA) enhanced via movable antenna (MA). We focuses on maximizing the effective throughput for a core user while ensuring reliable communication for an edge user by optimizing the MAs' coordinates and the power and rate allocations from the access point (AP). The optimization challenge is approached by decomposing it into two subproblems, utilizing successive convex approximation (SCA) to handle the highly non-concave nature of channel gains. Numerical results confirm that the proposed solution offers substantial improvements in effective throughput compared to NOMA short-packet communication with fixed position antennas (FPAs).
\end{abstract}


% \fi
\IEEEpeerreviewmaketitle

\section{Introduction}
\IEEEPARstart{I}{n} the era of the Internet of Things (IoT), the necessity for low-latency communication has become increasingly pronounced, owing to its pivotal role across a spectrum of application domains \cite{hamidi20215g}. In industrial automation, seamless coordination and real-time responsiveness are paramount, low-latency communication ensures the smooth operation of machinery and equipment. Similarly, remote control systems heavily rely on low-latency communication to enable the effective operation of unmanned vehicles and remote-operated machinery from a distance.

Short-packet communication is recognized as an effective method to minimize communication latency, though it suffers from notable decoding errors with finite block lengths \cite{wu2021comprehensive}. Non-orthogonal multiple access (NOMA) improves wireless system efficiency and capacity by allowing multiple users to simultaneously share the same spectral resources \cite{energy2023Ihsan}, thus increasing connection density and reducing latency. NOMA differentially allocates power to users to ensure fairness and minimize transmission conflicts, making it suitable for integrating with short-packet communication to address the IoT's low-latency demands. \cite{sun2018short} introduced NOMA into short-packet communication, aiming to leverage the advantages to meet the low-latency requirements of the IoT era.

In recent years, movable antennas (MAs) have garnered significant attention due to their sufficient utilization of spatial degrees of freedom (DoFs) \cite{wang2024movable}, which further enhances the performance of wireless communication systems. \cite{qin2024antenna} introduces a fluid antenna-assisted multi-user downlink system, optimizing positioning and beamforming to enhance performance and minimize transmission power. \cite{zhou2024movable} introduces movable antennas in MISO NOMA downlink systems, optimizing power allocation and antenna positioning to improve channel capacity. The implementation, mathematical modeling, and performance analysis of communication systems utilizing MAs are comprehensively detailed and analyzed in \cite{zhu2023modeling}. Meanwhile, \cite{ma2023mimo} presented advanced multiple-input multiple-output (MIMO) system where MAs are adjusted to boost system capacity by reshaping the MIMO channel effectively. \cite{zhu2023movable} showed that adjusting MAs optimally can significantly reduce the total transmit power in MA-enhanced multiple access
systems by creating favorable channel conditions. Based on these works, we consider whether MA can be applied to NOMA short-packet communication via antenna position optimization to further amplify its advantages.

In this letter, we investigate a short-packet downlink transmission system with NOMA assisted by MAs, where an access point (AP) transmits information to two users, each equipped with an MA. Considering the decoding error probability due to finite block length, our objective is to maximize the effective rate of successful data reception for core user (CU) while ensuring reliable communication for edge user (EU), by optimizing the coordinates of the MAs and the power and rate allocations of the AP. We decompose the complex non-convex optimization problem into two subproblems. Using successive convex approximation (SCA) technique, we obtain a local optimal solution for the first subproblem. Detailed steps for resolving the second subproblem are provided. Numerical results demonstrate that our proposed solution significantly enhances the effective rate of successful data reception for core user compared to NOMA short-packet communication with fixed position antennas (FPAs).

\section{System Model and Problem Formulation}

In Fig. 1, an AP transmits information to two users simultaneously within the same frequency band using finite block lengths, where AP is equipped with a FPA, while each user is equipped with a MA capable of movement within a square region of side length $A$. The coordinates of $u_{i}$'s MA can be represented as $\mathbf{u}_{i}=[x_{i},y_{i}]^{T},-\frac{A}{2} \leq x_{i},y_{i} \leq \frac{A}{2}$, where $i \in \{1,2\}$. Here, the user with better communication conditions, characterized by higher channel gains, is denoted as $u_{1}$, while the other user is denoted as $u_{2}$. We assume that the channel model adopted is based on the far-field response. This implies that although MA are deployed at the user terminal, the angle-of-departure (AoD), the angle-of-arrival (AoA), and the amplitude of the complex coefficient of each channel path remain unchanged \cite{zhu2023modeling}.

\begin{figure}[t]
\centering
\includegraphics[width=0.33\textwidth]{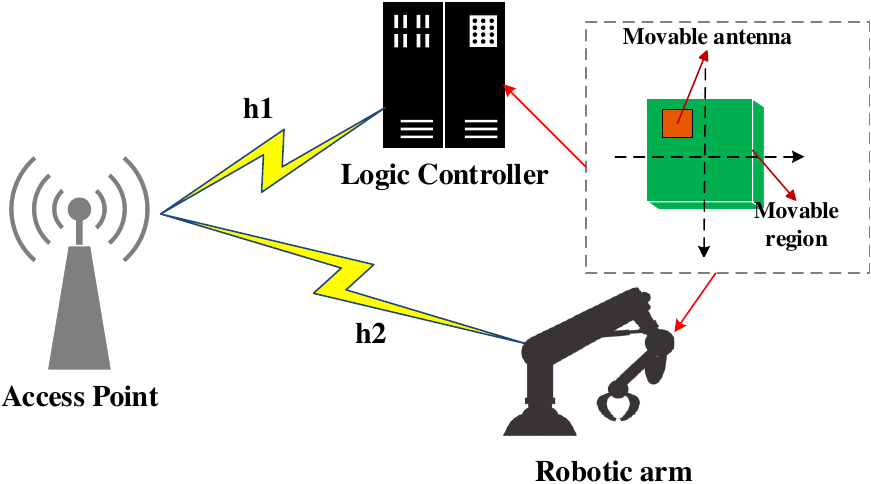}
\caption{System model diagram. In an actual intelligent factory automation communication system, an AP, a logic controller, and a robotic arm can be considered the smallest communication unit.}
\label{systemfigure}
\end{figure}

\subsection{Channel Modeling}
The mathematical expression for the signal received by the user $u_{i}$ is given by
\begin{align}
    y_{i}=h_{i} x+n_{i},i \in \{1,2\}
\end{align}
where $x$ is the input of the downlink transmission and ${h}_{i} \in \mathbb{N}$ is the channel coefficient from the AP to $u_{i}$. $x=\sqrt{P_{1}}x_{1}+\sqrt{P_{2}}x_{2}$, where $x_{1} \in \mathbb{C}$ and $x_{2} \in \mathbb{C}$ represent the data symbols with unit power transmitted to two users. $P_{1}$ and $P_{2}$ represent the transmit powers allocated by the AP to $u_{1}$ and $u_{2}$, respectively. The term $n_{i}$ originates from the influence of the additive white Gaussian noise (AWGN) channel, $n_{i} \sim \mathcal{CN}(0,\sigma_{i}^{2})$. $h_{i}$ can be further decomposed into a product of several factors, as follows:
\begin{align}
    h_{i}= \mathbf{w}_{i}^{H} \mathbf{f}_{i}(\mathbf{u}_{i}).
\end{align}
The interpretation of the product term mentioned in the formula is as follows:
\begin{itemize}
    \item $\mathbf{w}_i \in \mathbb{C}^{L_{i,r} \times 1}$ embodies the effective path response vector (EPRV) of $u_i$ within the receive region.
    \item $\mathbf{f}_{i}(\mathbf{u}_{i}) ={[e^{j \frac{2\pi}{\lambda} \mathbf{u}_{i}^{T} \mathbf{\rho}_{i,1}},\ldots,e^{j \frac{2\pi}{\lambda} \mathbf{u}_{i}^{T} \mathbf{\rho}_{i,L_{i,r}}}]}^{T} \in \mathbb{C}^{L_{i,r} \times 1}$ is the field response vector (FRV) in the receive region of $u_{i}$, where the definition $\mathbf{\rho}_{i,k} \triangleq {[\sin{\theta_{i,k}}\cos{\phi_{i,k}},\cos{\theta_{i,k}}]}^{T}$ holds. The elevation and azimuth angles of the $u_{i}$'s $k$-th receive path are denoted by $\theta_{i,k}$ and $\phi_{i,k}$, respectively, for $k=1,2,...,L_{i,r}$. Furthermore, $\lambda$ is the wavelength.
\end{itemize}

\subsection{Short-Packet NOMA Transmission}
Applying successive interference cancellation (SIC) at the receiver to recover signals is an important aspect of NOMA \cite{cover1972broadcast}. Here, we assume that the upper bound of two users' channel gains satisfies the following relationship:
\begin{align}
    {\left|\left| \mathbf{w}_{1} \right|\right|}_1 > {\left|\left| \mathbf{w}_{2} \right|\right|}_1,
\end{align}
where the notation ${\left|\left| \cdot \right|\right|}_1$ denotes the $l_1$ norm. Taking fairness into consideration, NOMA strategies allocate a greater amount of power for communication with $u_{2}$. Owing to this premise, $u_{1}$ can eliminate interference from $u_{2}$ through SIC. However, $u_{2}$ cannot eliminate interference from $u_{1}$ through SIC.

In short-packet communication, the decoding error probability at the receiver is non-negligible, primarily due to the finite block length \cite{sun2018short}. We designate the effective decoding error probability at $u_{i}$ as $\epsilon_{i}$. According to \cite{polyanskiy2010channel}, the decoding error probability of $x_{2}$ at $u_{1}$ for a given transmission rate $R_{2}$ determined by AP, denoted by $\epsilon_{2}^{1}$, can be approximated by
\begin{align}
    \epsilon_{2}^{1} \approx Q(f(\gamma_{2}^{1},N,R_{2})),
\label{dep 12}
\end{align}
%where the Q-function is defined as $Q(x) \triangleq \frac{1}{\sqrt{2\pi}}\int_{x}^{\infty}\exp{(-\frac{t^{2}}{2})}dt$. The computational expression for the $f(\cdot)$, which has three independent variables, is given by $f(x,y,z)=\ln{2}\sqrt{\frac{y}{1-{(1+x)}^{-2}}}(\log_{2}(1+x)-z)$, where $N$ is the block length and $\gamma_{2}^{1}$ is the signal-to-inference-plus-noise (SINR) of $x_{2}$ at $u_{1}$. In particular, $\gamma_{2}^{1}$ is given by
where Q-function is defined as $Q(x) \triangleq \frac{1}{\sqrt{2\pi}}\int_{x}^{\infty}\exp{(-\frac{t^{2}}{2})}dt$. The computational expression for $f(\cdot)$ can be given by $f(x,y,z)=\ln{2}\sqrt{\frac{y}{1-{(1+x)}^{-2}}}(\log_{2}(1+x)-z)$, where $N$ is the block length and $\gamma_{2}^{1}$ is the signal-to-inference-plus-noise (SINR) of $x_{2}$ at $u_{1}$. In particular, $\gamma_{2}^{1}$ is given by
\begin{align}
    \gamma_{2}^{1}=\frac{{\left| h_{1} \right|}^{2}P_{2}}{{\left| h_{1} \right|}^{2}P_{1}+\sigma_{1}^{2}}. \label{SINR 12}
\end{align}
Providing that SIC succeeds, the SNR of $s_{1}$ at $u_{1}$ and the decoding error probability of $x_{1}$ at $u_{1}$ can be written as
\begin{align}
    \gamma_{1}^{1}=\frac{{\left| h_{1} \right|}^{2}P_{1}}{{\sigma_{1}}^2}, \label{SNR 11} \\
    \epsilon_{1}^{1} \approx Q(f(\gamma_{1}^{1},N,R_{1})).
\label{dep 11}
\end{align}
If SIC does not succeeds, the SINR of $s_{1}$ at $u_{1}$ and the decoding error probability of $x_{1}$ at $u_{1}$ can be calculated as
\begin{align}
    {\gamma_{1}^{1}}'=\frac{{\left| h_{1} \right|}^{2}P_{1}}{{\left| h_{1} \right|}^{2}P_{2}+\sigma_{1}^{2}}, \label{SINR11} \\
    {\epsilon_{1}^{1}}' \approx Q(f({\gamma_{1}^{1}}',N,R_{1})),
\label{dep 11'}
\end{align}
where $R_{1}$ is the transmission rate determined by AP. The effective decoding error probability at $u_{1}$ is expressed by
\begin{align}
    \epsilon_{1}=\left \{
    \begin{array}{l}
    \epsilon_{1}^{1}(1-\epsilon_{2}^{1})+{\epsilon_{1}^{1}}'\epsilon_{2}^{1}, \text{ } R_{1} \leq \log_{2}(1+{\gamma_{1}^{1}}'), \\
    \epsilon_{1}^{1}(1-\epsilon_{2}^{1})+\epsilon_{2}^{1}, \text{ } \log_{2}(1+{\gamma_{1}^{1}}') \leq R_{1} \leq \log_{2}(1+\gamma_{1}^{1}).
    \end{array}
    \right.
\label{dep 1}
\end{align}
The signal-to-inference-noise (SINR) of at $u_{2}$ can be given by
\begin{align}
    \gamma_{2}^{2}=\frac{{\left| h_{2} \right|}^{2}P_{2}}{{\left| h_{2} \right|}^{2}P_{1}+\sigma_{2}^{2}}. \label{SINR 22}
\end{align}
Then the effective decoding error probability of directly detecting $x_{2}$ at $u_{2}$ is given by
\begin{align}
    \epsilon_{2} \approx Q(f(\gamma_{2}^{2},N,R_{2})).
\label{dep 2}
\end{align}

Due to short-packet communication strategies, zero decoding error probability is unattainable, so effective throughput (bits per channel use) is used as the performance metric \cite{sun2018short}. The effective throughput $T_{i}$ for user $u_{i}$ is given by
\begin{align}
    T_{i}=\frac{N_{i}}{N}R_{i}(1-\epsilon_{i}),
\end{align}
where $N_{i}$ represents the block length assigned to $u_{i}$ and $R_{i}$ signifies the transmission rate for $u_{i}$ within the finite block length regime. Due to the absence of time division multiple access (TDMA) in NOMA transmission, it follows that $N_{i}=N$.

\subsection{Problem Formulation}
We aim to maximize the effective throughput of $u_{1}$ while constraining the maximum allocated power at the AP and ensuring communication reliability for $u_{2}$ (i.e., guaranteeing that the effective throughput of $u_{2}$ exceeds a certain threshold). Mathematically, the formulated optimization problem can be expressed as
\begin{subequations}
    \begin{align}
    \max_{\{R_{1},R_{2},P_{1},P_{2},\mathbf{u}_{1},\mathbf{u}_{2}\}} \quad & T_{1}, \\
    \text{subject to} \quad & P_{1}+P_{2} \leq P_{max}, \\
    & T_{2} \geq T_{0}, \label{constraint b} \\
    & \mathbf{u}_{1},\mathbf{u}_{2}
        \in
        \mathcal{C},
\end{align}
\label{Optimization problem}
\end{subequations}
where $\mathcal{C}=\{ {[x_{1},x_{2}]}^{T}|-\frac{A}{2}<x_{1},x_{2}<\frac{A}{2} \}$. $P_{max}$ denotes the average transmit power allocated by the AP within a block. $T_{0}$ represents the minimum effective throughput required to achieve reliable transmission for $u_{2}$.

\section{The Proposed Solution}\label{section 3}
Since optimization variables $P_1, P_2, R_1$, and $R_2$ are independent of each other. In (\ref{Optimization problem}), the original optimization problem can be decomposed into two sets of optimization variables.
To be specific, we can provide an expression for the channel gain, which is determined by the coordinates of the MAs as follows:
\begin{align}
    {\left| h_{i} \right|}^{2} &= \mathbf{f}_{i}^{H} (\mathbf{u}_{i}) \mathbf{w}_{i} \mathbf{w}_{i}^{H} \mathbf{f}_{i}(\mathbf{u}_{i}) \nonumber \\
    &\triangleq \mathbf{f}_{i}^{H} (\mathbf{u}_{i}) \mathbf{W}_{i} \mathbf{f}_{i}(\mathbf{u}_{i}), i=1,2
\end{align}
where $\mathbf{W}_{i} \triangleq \mathbf{w}_{i} \mathbf{w}_{i}^{H} \in \mathbb{C}^{L_{i,r} \times L_{i,r}}$ denotes a constant positive semi-definite matrix. Since ${\left| h_{i} \right|}^{2}$ is highly non-concave with respect to $\mathbf{u}_{i}$ and forming the feasible set from constraints ($\ref{constraint b}$) is complicated, directly addressing Problem (\ref{Optimization problem}) poses considerable difficulty. Fortunately, channel gains for two users depend solely on the MA coordinates, affecting their own effective throughput individually. Our strategy entails decomposing the problem into two subproblems, which are more amenable to resolution, leveraging Proposition 1.

\begin{proposition}
    When other variables are given, the effective throughputs $T_{1}$ and $T_{2}$ are respectively monotonically increasing functions of the channel gain ${\left|h_{1}\right|}^{2}$ and ${\left|h_{2}\right|}^{2}$.
\end{proposition}

\emph{Proof:} The detailed proof is provided in Appendix.
     $\hfill\blacksquare$

Subproblem 1: Seek the optimal coordinates of the MAs to maximize the channel gain for users.
\begin{subequations}
    \begin{align}
        \max_{\{ \mathbf{u}_{1},\mathbf{u}_{2} \}} \quad & {\left| h_{i} \right|}^{2} = \mathbf{f}_{i}^{H} (\mathbf{u}_{i}) \mathbf{W}_{i} \mathbf{f}_{i}(\mathbf{u}_{i}), \\
        \text{subject to} \quad & \mathbf{u}_i \in \mathcal{C},\forall i \in \{1,2\}, \label{sp1 constraint}
    \end{align}
    \label{subproblem 1}
\end{subequations}
Subproblem 2: Seek the optimal resource allocation strategy at the AP under the condition of maximizing channel gain.
\begin{subequations}
    \begin{align}
        \max_{\{R_{1},R_{2},P_{1},P_{2}\}} \quad & T_{1}, \\
        \text{subject to} \quad & P_{1}+P_{2} \leq P_{max}, \\
        & T_{2} \geq T_{0}, \label{sp2c2}
    \end{align}
    \label{subproblem 2}
\end{subequations}
Subsequently, we will proceed to solve these two subproblems in the remaining part of Section \ref{section 3}.

\subsection{Solution to Subproblem 1}
We employ SCA technique to address Subproblem 1, thereby obtaining the optimal values of the coordinates of the MAs \cite{ma2023mimo}. Given any point $\mathbf{u}_{i}^{k}$ within the feasible set, $\mathbf{f}_{i}(\mathbf{u}_{i}^{k})$ is determined. Here, $\mathbf{u}_{i}^{k}$ can serve as the point given in the $k$-th iteration of SCA. Then, the global lower bound can be provided by the following inequality:
\begin{align}
    {\left| h_{i} \right|}^{2} \geq & \mathbf{f}_{i}^{H} (\mathbf{u}_{i}^{k}) \mathbf{W}_{i} \mathbf{f}_{i}(\mathbf{u}_{i}^{k}) \nonumber \\
    & + 2\text{Re}\{ \mathbf{f}_{i}^{H}(\mathbf{u}_{i}^{k}) \mathbf{W}_{i} (\mathbf{f}_{i}(\mathbf{u}_{i})-\mathbf{f}_{i}(\mathbf{u}_{i}^{k})) \} \nonumber \\
    = & 2\text{Re}\{ \mathbf{f}_{i}^{H}(\mathbf{u}_{i}^{k}) \mathbf{W}_{i} \mathbf{f}_{i}(\mathbf{u}_{i}) \} \nonumber \\
    & - \mathbf{f}_{i}^{H} (\mathbf{u}_{i}^{k}) \mathbf{W}_{i} \mathbf{f}_{i}(\mathbf{u}_{i}^{k}). \label{lb 1}
\end{align}
Without loss of generality, let us define $G(\mathbf{u}_{i}) \triangleq \text{Re}\{ \mathbf{f}_{i}^{H}(\mathbf{u}_{i}^{k}) \mathbf{W}_{i} \mathbf{f}_{i}(\mathbf{u}_{i}) \} \label{lb 1}$.

After the aforementioned operations, $G(\mathbf{u}_{i})$ is still not a concave function with respect to $\mathbf{u}_{i}$. Therefore, considering the ease of analysis when using a quadratic function as a surrogate to approximate the lower bound of the original function, we perform a second-order Taylor expansion of $G(\mathbf{u}_{i})$ at $\mathbf{u}_{i}^{k}$, thereby obtaining a quadratic surrogate lower bound for it \cite{ma2023mimo}:
\begin{align}
    G(\mathbf{u}_{i}) \geq & G(\mathbf{u}_{i}^{k})+\nabla^{T} G(\mathbf{u}_{i}^{k}) (\mathbf{u}_{i}-\mathbf{u}_{i}^{k}) \nonumber \\
    &-\frac{\delta_{i}^{k}}{2}(\mathbf{u}_{i}-\mathbf{u}_{i}^{k})^{T}(\mathbf{u}_{i}-\mathbf{u}_{i}^{k}) \nonumber \\
    = & \underbrace{ -\frac{\delta_{i}^{k}}{2} \mathbf{u}_{i}^{T}\mathbf{u}_{i} + (\nabla G(\mathbf{u}_{i}^{k})+\delta_{i}^{k}\mathbf{u}_{i}^{k})^{T}\mathbf{u}_{i} \label{qslb}}_{\overline{G}(\mathbf{u}_{i})} \nonumber \\
    &+ \underbrace{ G(\mathbf{u}_{i}^{k})-(\nabla G(\mathbf{u}_{i}^{k})+\frac{\delta_{i}^{k}}{2}\mathbf{u}_{i}^{k})^{T}\mathbf{u}_{i}^{k}}_{\text{constant}}.
\end{align}
where $\nabla G(\mathbf{u}_{i})$ denotes the gradient vector of $G(\mathbf{u}_{i})$ over $\mathbf{u}_{i}$ and $\delta_{i}^{k}$ is a positive number. After the aforementioned analysis, Subproblem 1 can be formulated as maximizing $\overline{G}(\mathbf{u}_{i})$. The value of $\delta_{i}^{k}$ and the expression for $\nabla G(\mathbf{u}_{i})$ are provided in the Appendix \ref{appendix b}. Due to the quadratic nature of $\overline{G}(\mathbf{u}_{i})$ and the fact that the solution of the quadratic function can be expressed in closed form, the optimization problem for the $k$-th iteration of SCA can be given by
\begin{subequations}
    \begin{align}
        \max_{\mathbf{u}_{1},\mathbf{u}_{2}} \quad & \overline{G}(\mathbf{u}_{i}), \\
        \text{subject to} \quad & \mathbf{u}_{i} \in \mathcal{C},\forall i \in \{1,2\}.
    \end{align}
    \label{i-th problem}
\end{subequations} \\
The optimal $\mathbf{u}_{i}^{k+1,*}$ obtained from solving optimization problem (\ref{i-th problem}) in the $k$-th iteration will be utilized for the subsequent iteration of SCA.

\subsection{Solution to Subproblem 2}
Here, we address Problem (17) as follows. Firstly, the equality $P_{1}+P_{2}=P_{max}$ is always guaranteed when maximizing $T_{1}$ subject to $T_{2} \geq T_{0}$. Secondly, the equality $T_{2}=T_{0}$ is always guaranteed when maximizing $T_{1}$. Next, we provide an overview of the steps for solving Subproblem 2:
\begin{itemize}
    \item \textit{Step 1}: Given a feasible $P_2$, since the value of $T_2$ is independent of $R_1$ and $P_1$ can be expressed as $P_{max}-P_2$, we can utilize the equation $T_2=T_0$ to solve for the desired $R_2$. Here, a fixed-point iteration algorithm will be employed for the solution of $R_2$ (i.e., $R_2^*$):
    \begin{align}
        R_{2}:=\frac{T_{0}}{1-Q(f(\gamma_{2}^{2},N,R_{2}))}.
    \end{align}
    To enforce Constraint (\ref{sp2c2}), $P_2$ is subject to a lower bound, which will be provided in the third step.
    \item \textit{Step 2}: Now that $P_{1}$, $P_{2}$, and $R_{2}$ are all determined, we proceed to seek $R_{1}$ that maximizes $T_{1}$. It can be given by
    \begin{align}
        R_{1}^{*}=\left\{
        \begin{aligned}
        R_{1}^{\dagger},& \text{if } 0 \leq R_{1} \leq \log_{2}(1+{\gamma_{1}^{1}}'), \\
        R_{1}^{\ddagger},& \text{if } \log_{2}(1+{\gamma_{1}^{1}}') \leq R_{1} \leq \log_{2}(1+\gamma_{1}^{1}).
        \end{aligned}
        \right.
    \end{align}
    The values $R_{1}^{\dagger}$ and $R_{1}^{\ddagger}$ represent the roots of the equations $(1-\epsilon_{2}^{1})\mathscr{U}(\gamma_{1}^{1},N,R_{1}^{\dagger})+\epsilon_{2}^{1}\mathscr{U}({\gamma_{1}^{1}}',N,R_{1}^{\dagger})=0$ and $\mathscr{U}(\gamma_{1}^{1},N,R_{1}^{\ddagger})=0$ respectively. $\mathscr{U}(x,y,z)$ is defined as
    \begin{align}
        \mathscr{U}(x,y,z) \triangleq & 1-Q(f(x,y,z)) \nonumber \\
        &-\frac{yz\ln{2}}{\sqrt{2\pi(1-{(1+x)}^{-2})}}e^{-\frac{f^{2}(x,y,z)}{2}}.
    \label{u_function}
    \end{align}
    \item \textit{Step 3}: The content of the third step mainly provides the lower bound for $P_2$. Assuming the lower bound for $P_2$ is $P_{2}^{l}$, then $P_{2}^{l}$ is the root of the following equation:
    \begin{align}
        R_{2}^{\dagger}(1-Q(f(\gamma_{2}^{l},N,R_{2}^{\dagger})))=T_{0}.
    \end{align}
    The value $R_{2}^{\dagger}$ represents the root of $\mathscr{U}(\gamma_{2}^{l},N,R_{2}^{\dagger})=0$ and $\gamma_{2}^{l}=P_{2}^{l} {\left| h_{2} \right|}^{2}/((P_{max}-P_{2}^{l}){\left| h_{2} \right|}^{2}+\sigma_{2}^{2})$.
    \item \textit{Step 4}: The feasible set of $P_{2}$ is $P_{2}^{l} \leq P_{2} \leq P_{max}$. Given that the feasible set of $P_{2}$ has been specified and is of finite length, we can employ linear search or golden section search to obtain the optimal value of $P_{2}$, denoted as $P_{2}^{*}$. Then the optimal value of $P_{1}$ is $P_{max}-P_{2}^{*}$.
\end{itemize}

% \begin{figure}[t]
%     \centering
%     \includegraphics[width=0.308\textwidth]{T1-P2 (2).pdf}
%     \caption{The curve representing the maximum achievable value of $T_1(\text{ bps/Hz})$ as $P_2(\text{dB})$ varies, where $T_{0}=1\text{ bps/Hz}$ and $N=100$.}
%     \label{fig1}
% \end{figure}

% \begin{figure}[t]
%     \centering
%     \includegraphics[width=0.32\textwidth]{T1-N (2).pdf}
%     \caption{This figure illustrates the curve of the maximum achievable values of $T_1(\text{ bps/Hz})$ for different values of $N$, where $T_{0}=2\text{ bps/Hz}$.}
%     \label{fig2}
% \end{figure}

% \begin{figure}[t]
%     \centering
%     \includegraphics[width=0.308\textwidth]{T1-T0 (2).pdf}
%     \caption{This figure illustrates the maximum achievable value of $T_1(\text{ bps/Hz})$ as T0 varies with the block length $N$ set to 200.}
%     \label{fig3}
% \end{figure}

\begin{figure*}[t]
\centering
\subfigure[Graph of $T_1$ versus $P_2$ for $N=100$ and $T_0=1$ bps/Hz.]{
\begin{minipage}[t]{0.308\textwidth}
\centering
\includegraphics[width=\textwidth]{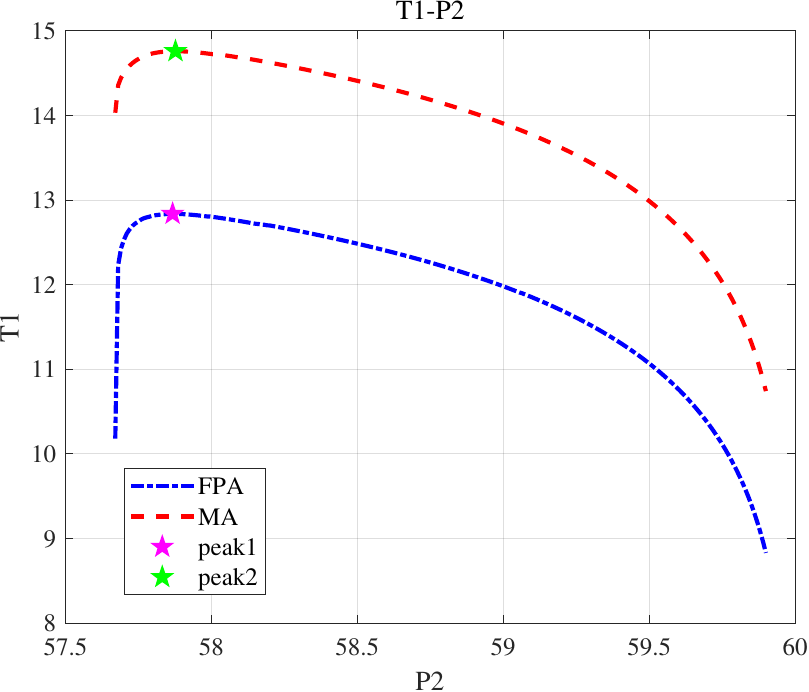}
% \caption{fig1}
\end{minipage}%
}%
\subfigure[Graph of $T_1$ versus $N$ for $T_0=2$ bps/Hz.]{
\begin{minipage}[t]{0.32\textwidth}
\centering
\includegraphics[width=\textwidth]{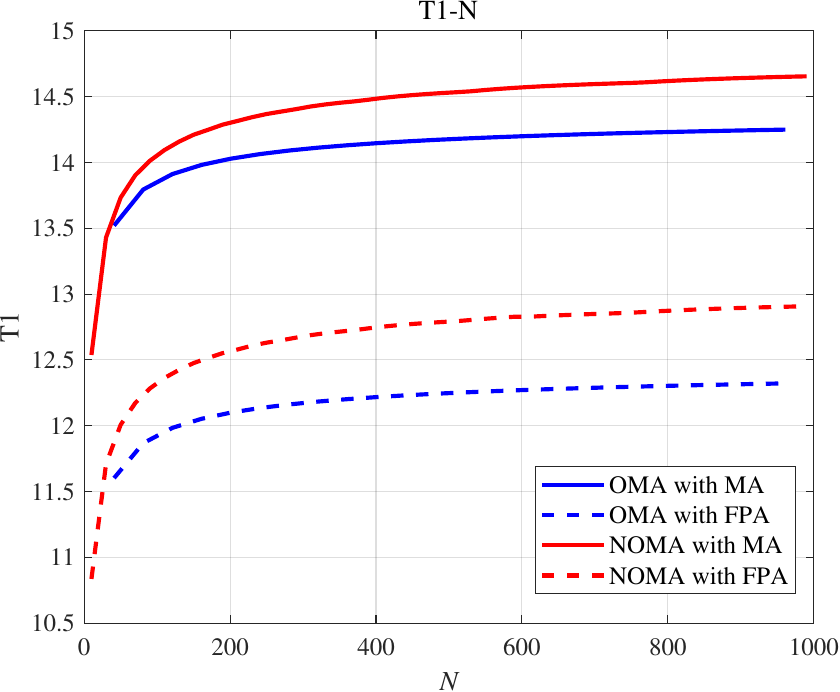}
% \caption{fig2}
\end{minipage}%
}%
\subfigure[Graph of $T_1$ versus $T_0$ for $N=200$.]{
\begin{minipage}[t]{0.308\textwidth}
\centering
\includegraphics[width=\textwidth]{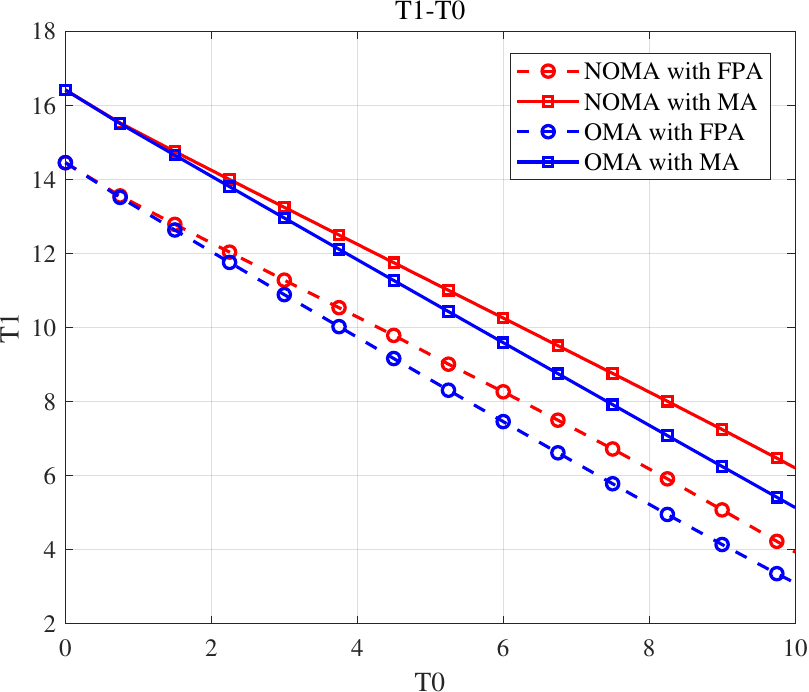}
% \caption{fig3}
\end{minipage}
}%
\caption{Simulation results}
\label{simufig}
\end{figure*}

\section{Numerical Results}
In this section, we present numerical results to validate the effectiveness of our MA-assisted short-packet downlink transmission with NOMA scheme by comparing it to a scheme where MA is replaced by FPA. The distances between $u_1$, $u_2$, and the AP are set to 20 m and 60 m, respectively. Each user's noise power $\delta_{i}^{2}$ is normalized, and the ratio of AP's average transmit power ($P_{max}$) to noise power is 60 dB. The movable region side length for MAs is $A=3\lambda$, with initial MA positions at the origin, i.e., ${\{ \mathbf{u}_{i}^{0} \}_{i=1}^{2}}=[0,0]^{T}$. Each user has 4 receive paths (${L_{i,r}}=4$). The EPRV of $u_i$ is modeled as a circular symmetric complex Gaussian random vector, with $n$-th $(1 \leq n \leq L_{i,r})$ component $w_{i,n} \sim \mathcal{C}\mathcal{N}(0,d_{i}^{-\alpha}/L_{i,r})$, where $\alpha=1$ and $d_{i}$ is the distance from $u_i$ to the AP. Elevation and azimuth AoDs of each user are modeled as independent and identically distributed uniform distributions over $[0,\pi]$. The simulation results presented in this section were obtained by aggregating the outcomes of simulations with repetition counts on the order of $10^3$ and subsequently computing their average.

%We denote by $\eta_{i}$ the ratio of the optimized channel gain of $u_{i}$ to its value before optimization, i.e., $\eta_i=\mathbf{f}_{i}^{H} (\mathbf{u}_{i}^{*}) \mathbf{W}_{i} \mathbf{f}_{i}(\mathbf{u}_{i}^{*})/\mathbf{f}_{i}^{H} (\mathbf{u}_{i}^{0}) \mathbf{W}_{i} \mathbf{f}_{i}(\mathbf{u}_{i}^{0})$, where $\mathbf{u}_{i}^{*}$ and $\mathbf{u}_{i}^{0}$ represent the optimal solution coordinates and initial coordinates of MA for $u_i$ in each experiment. 

In Fig. \ref{simufig}(a), We plotted the relationship between $T_1$ and $P_2$, comparing our proposed scheme with the benchmark scheme, while optimizing the corresponding parameters (e.g., $P_1$,$R_1$,$R_2$) accordingly. In this figure, we first observe that our proposed scheme consistently outperforms the benchmark scheme in optimizing the effective throughput of $u_1$ regardless of the value of $P_2$. We can find that the introduction of MA led to an approximately $16\%$ improvement in the effective throughput of $u_1$ compared to the benchmark scheme. In fact, the maximum effective throughput of $u_1$ under the benchmark scheme serves as a lower bound for that under our proposed scheme. %Furthermore, we observe that our proposed scheme does not significantly alter power allocation when maximizing $T_1$, as the numerical value of power far exceeds that of the channel gain. Consequently, even with an increased channel gain, the relative difference remains substantial. Thus, when optimizing $T_1$, only a minor increase in power is required to balance the enhanced channel gain.

In Fig. \ref{simufig}(b), we show that our proposed scheme achieves superior low-latency transmission by optimizing the effective throughput of $u_1$ for varying block lengths, compared to other schemes. With symbol duration $T_s$, the delay for a block length $N$ is $NT_s$, indicating longer block lengths increase delay. Our scheme consistently outperforms the benchmark regardless of $N$. Notably, to achieve 13 bps/Hz, the benchmark requires nearly infinite block lengths, while our scheme achieves this with a block length under 100. In addition, within the experimental communication scenario, the superiority of NOMA over OMA in achieving low latency has been verified. Regardless of whether the users are equipped with MAs or FPAs, when the block length is fixed, the maximum throughput achieved by $u_1$ employing NOMA is consistently higher than that achieved using OMA.

Fig. \ref{simufig}(c) depicts the variation of $T_1$ with respect to the constraint $T_0$. Firstly, $T_1$ decreases as $T_0$ increases across different schemes. Secondly, the introduction of MA significantly enhances $u_1$'s effective throughput in both NOMA and OMA short-packet communications. Finally, when $T_0$ and $T_1$ are comparable in value, NOMA short-packet communication outperforms OMA in maximizing T1, with a more pronounced difference between the two.

\section{Conclusion}
In this letter, we investigated the MA-enhanced NOMA short-packet transmission performance for the downlink communication system. We focused on maximizing the effective throughout for the CU while ensuring the reliability of transmissions to the EU. This was done under the conditions of limiting the AP's average transmission power within a block, securing a minimum effective throughput for the EU, and confining the movable region of the MA for each user. Numerical results confirm that our proposed approach markedly improves the effective throughout for the CU, outperforming traditional NOMA short-packet communication systems with FPA.

% or


\appendices
\section{Proof of Proposition 1}

Combining (\ref{dep 1}), the effective throughputs $T_{1}$ and $T_{2}$ in (\ref{Optimization problem}) can be further expressed by the following equations:
\begin{align}
    T_{1}&=R_{1}(1-\epsilon_{1}^{1}+(\epsilon_{1}^{1}-I)\epsilon_{2}^{1}), \label{T1} \\
    T_{2}&=R_{2}(1-\epsilon_{2}), \label{T2}
\end{align}
where $I$ depends on the value of $R_{1}$. If $R_{1} \leq \log_{2}(1+{\gamma_{1}^{1}}')$ holds, $I={\epsilon_{1}^{1}}'$, and if $\log_{2}(1+{\gamma_{1}^{1}}') \leq R_{1} \leq \log_{2}(1+\gamma_{1}^{1})$ holds, $I=1$. \iffalse As equations (\ref{dep 12}), (\ref{dep 11}), (\ref{dep 11'}) and (\ref{dep 2}) all involve the function form $Q(f(x,y,z))$, we proceed to analyze the monotonicity of $Q(f(x,y,z))$ with respect to $x$.\fi According to \cite{sun2018short}, $Q(f(x,y,z))$ is a monotonically decreasing function with respect to $x$. Then, following (\ref{SNR 11}) and (\ref{SINR11}) we have $\gamma_{1}^{1}>{\gamma_{1}^{1}}'$, which leads to $\epsilon_{1}^{1}<{\epsilon_{1}^{1}}' \leq 1$ as per (\ref{dep 11}) and (\ref{dep 11'}). Considering a function 
$g(x)=\frac{ax}{bx+c}$ with domain $x>0$, where $a$, $b$, and $c$ are all constants greater than zero, and its derivative with respect to $x$ is $g'(x)=\frac{ac}{(bx+c)^{2}}$, it is evident that $g'(x)$ is greater than zero. Thus, $g(x)$ is a monotonically increasing function with respect to $x$. This implies that the SNR/SINR will increase as the channel gain increases as per (\ref{SINR 12}), (\ref{SNR 11}), (\ref{SINR11}) and (\ref{SINR 22}).

%Based on the aforementioned conclusion, an increase in ${\left| h_{1} \right|}^{2}$ will result in the increase of $\gamma_{2}^{1}$ and $\gamma_{1}^{1}$, consequently leading to the decrease of $\epsilon_{2}^{1}$ and $\epsilon_{1}^{1}$. Combined with the previously proven $\epsilon_{1}^{1}-{\epsilon_{1}^{1}}'<0$ and $\epsilon_{1}^{1}-1<0$, it can be inferred that $\frac{\partial T_{1}}{\partial \epsilon_{2}^{1}}<0$, implying that $T_{1}$ increases with the decrease of $\epsilon_{2}^{1}$ and $\epsilon_{1}^{1}$. The proof of the monotonicity of $T_{2}$ with respect to ${\left| h_{2} \right|}^{2}$ is simpler. An increase in ${\left| h_{2} \right|}^{2}$ leads to an increase in $\gamma_{2}^{2}$, further resulting in a decrease in $\epsilon_{2}$, and ultimately leading to an increase in $T_{2}$ as per (\ref{T2}).

Based on this, an increase in ${\left| h_{1} \right|}^{2}$ leads to increases in $\gamma_{2}^{1}$ and $\gamma_{1}^{1}$, resulting in decreases in $\epsilon_{2}^{1}$ and $\epsilon_{1}^{1}$. Given that $\epsilon_{1}^{1}-I<0$, we infer that $\frac{\partial T_{1}}{\partial \epsilon_{2}^{1}}<0$, implying $T_{1}$ increases as $\epsilon_{2}^{1}$ and $\epsilon_{1}^{1}$ decrease. Similarly, an increase in ${\left| h_{2} \right|}^{2}$ raises $\gamma_{2}^{2}$, decreases $\epsilon_{2}$, and thus increases $T_{2}$ as per (\ref{T2}).

\label{appendix a}
\section{Expression for $\nabla G(\mathbf{u}_{i})$ and Value of $\delta_{i}^{k}$}
In the $k$-th iteration of SCA, it is indispensable to compute the gradient vector of $G(\mathbf{u}_{i})$ and $\delta_{i}^{k}$. For analytical convenience, we introduce the following definitions:
\begin{align}
    \mathbf{v} & \triangleq \mathbf{W}_{i} \mathbf{f}_{i}(\mathbf{u}_{i}^{k}) \\
    & = [b_{1} , \ldots , b_{l}]^{T} \nonumber = [\left| b_{1} \right|e^{j \angle b_{1}} , \ldots , \left| b_{l} \right|e^{j \angle b_{l}}]^{T} \in \mathbb{C}^{l \times 1}, \nonumber
\end{align}
where $l=L_{i,r}$. Then $G(\mathbf{u}_{i})$ can be further expressed as
\begin{align}
    &G(\mathbf{u}_{i})  = \text{Re} \{\mathbf{v}^{H} \mathbf{f}_{i}(\mathbf{u}_{i})\} \\
    & = \text{Re} \left\{ \sum\nolimits_{p=1}^{L_{i,r}} \left| b_{p} \right| e^{j[\frac{2 \pi}{\lambda}(x_{i} \sin{\theta_{i,p}} \cos{\phi_{i,p}}+y_{i} \cos{\theta_{i,p}})-\angle b_{p}]}\right\} \nonumber \\
    & = \sum\nolimits_{p=1}^{L_{i,r}} \left| b_{p} \right| \cos{(\Gamma_{p}(\mathbf{u}_{i}))}, \nonumber
\end{align}
where $\Gamma_{p}(\mathbf{u}_{i}) \triangleq \frac{2 \pi}{\lambda}(x_{i} \sin{\theta_{i,p}} \cos{\phi_{i,p}}+y_{i} \cos{\theta_{i,p}})-\angle b_{p}$. Subsequently, we can provide the expression for the gradient vector of $G(\mathbf{u}_{i})$ with respect to $\mathbf{u}_{i}$:
\begin{align}
    \nabla G(\mathbf{u}_{i})
    & = \left[\frac{\partial G(\mathbf{u}_{i})}{\partial x_{i}}, \frac{\partial G(\mathbf{u}_{i})}{\partial y_{i}}\right]^{T} \\
    & = 
    \begin{bmatrix}
        -\frac{2\pi}{\lambda} \sum_{p=1}^{L_{i,r}} \left| b_{p} \right| \sin{\theta_{i,p}} \cos{\phi_{i,p}} \sin{(\Gamma_{p}(\mathbf{u}_{i}))} \\
        -\frac{2\pi}{\lambda} \sum_{p=1}^{L_{i,r}} \left| b_{p} \right| \cos{\theta_{i,p}} \sin{(\Gamma_{p}(\mathbf{u}_{i}))} \nonumber
    \end{bmatrix}
\end{align}
According to \cite{ma2023mimo}, we have $\delta_{i}^{k} \mathbf{I}_{2} \succ \nabla^{2} G(\mathbf{u}_{i})$.
After some manipulations, we have
$\delta_{i}^{k}=\frac{8 \pi^{2}}{\lambda^{2}}\sum\nolimits_{p=1}^{L_{i,r}} \left| b_{p} \right|.$

\label{appendix b}

\ifCLASSOPTIONcaptionsoff
  \newpage
\fi



%\bibliographystyle{IEEEtran}
\bibliographystyle{ieeetr}
\bibliography{draft}

\end{document}